# Analytical, numerical, and experimental investigation of a Luneburg lens system for directional cloaking

C. Babayiğit,[1,*] Aydın S. Evren,[2] E. Bor,[1,2] H. Kurt,[1] and M. Turduev[2]

[1]*Department of Electrical and Electronics Engineering, TOBB University of Economics and Technology, Ankara 06560, Turkey*
[2]*Department of Electrical and Electronics Engineering, TED University, Ankara 06420, Turkey*

In this study, the design of a directional cloaking based on the Luneburg lens system is proposed and its operating principle is experimentally verified. The cloaking concept is analytically investigated via geometrical optics and numerically realized with the help of the finite-difference time-domain method. In order to benefit from its unique focusing and/or collimating characteristics of light, the Luneburg lens is used. We show that by the proper combination of Luneburg lenses in an array form, incident light bypasses the region between junctions of the lenses, i.e., the "dark zone." Hence, direct interaction of an object with propagating light is prevented if one places the object to be cloaked inside that dark zone. This effect is used for hiding an object which is made of a perfectly electric conductor material. In order to design an implementable cloaking device, the Luneburg lens is discretized into a photonic crystal structure having gradually varying air cylindrical holes in a dielectric material by using Maxwell Garnett effective medium approximations. Experimental verifications of the designed cloaking structure are performed at microwave frequencies of around 8 GHz. The proposed structure is fabricated by three-dimensional printing of dielectric polylactide material and a brass metallic alloy is utilized in place of the perfectly electric conductor material in microwave experiments. Good agreement between numerical and experimental results is found.

## I. INTRODUCTION

In the field of optics, realization of the optical cloaking effect has been an attractive topic that was initiated by the pioneering works based on transformation optics (TO) [1–3]. Here, TO simply bends the coordinate system to efficiently manipulate the direction of light propagation [4–7]. Since optical cloaking can be defined as the concealment of an object from an incident wave by bending and stretching the coordinate structures, it is not a feasible procedure because it requires unnatural materials with anisotropic, spatially varying permittivity and permeability values [8]. Nevertheless, metamaterials can provide this feature [9], but the realization of these structures is a challenging issue due to dimensions much smaller than the wavelength. In addition, the cloaking effect obtained by using the TO approach is narrowband and inherently lossy which is undesirable for optical applications. Hence, these listed difficulties force researchers to look for alternative solutions for the cloaking phenomenon.

In this regard, along with TO, several methods have been introduced to obtain the optical invisibility effect. Carpet cloaking, also known as the ground-plane cloak, is an approach for hiding objects under a specified refractive index layer made of isotropic, low-loss, and dielectric material [10–14]. Another approach is suppression of scatterings due to cloaked objects by using generalized Hilbert transforms to adapt the scattering response of the hiding object [15,16]. Moreover, Kramers-Kronig relations have been presented to minimize backscattering of waves which leads to optical cloaking [17]. Recently, an interesting approach based on metamaterials and optical neutrality was proposed to provide invisibility without evoking cloaking [18]. Last, but not least, the idea of using the optimization approach to obtain the cloaking effect shows promising results [19–22]. Here, the optimization methods search for possible designs of cloaking structures in accordance with a specific objective function. Furthermore, experimental verifications at the microwave frequency regime of cloaking designs based on optimization methods were reported in Refs. [23,24].

In addition to these studies, the focusing effect is also used for creating invisible regions both in ray and wave optics [25–27]. Moreover, the graded index (GRIN) optics can be considered as a powerful tool for efficient light manipulation. The GRIN medium efficiently bends the light to follow curved trajectories because of the gradual change in refractive index along the radial or axial directions [28–30]. For this reason, the GRIN medium provides an opportunity to obtain curved light without curved interfaces for the optical phenomena such as focusing and/or collecting, and diverging and/or spreading [31]. In this regard, conceptual studies on optical cloaking by using GRIN optics have been reported recently in [32–34].

In this study, we proposed the idea of using the GRIN optics concept to cloak a highly scattering cylindrical object made of a perfectly electric conductor (PEC) material. In particular, the combination of GRIN Luneburg lenses is used for the optical hiding purpose. The ray transfer analysis of Luneburg lenses as a cloaking system is analytically derived via geometrical optics. Next, in order to design a realizable cloaking device, the continuous GRIN Luneburg lens is discretized as a GRIN photonic crystal (PC) structure having varying radii of air cylindrical holes in a dielectric slab by using Maxwell Garnett approximations, i.e., effective medium theory (EMT). As a dielectric host material, polylactide (PLA) thermoplastic having effective permittivity of $\varepsilon_{PLA} = 2.4025$ is employed. The numerical analysis of the optical cloaking effect is conducted by the three-dimensional (3D) finite-difference time-domain (FDTD) method. Moreover, the designed GRIN PC Luneburg cloaking system is fabricated by using 3D printing technology and experimental verification of numerical results is performed at microwave frequencies. It should be noted that the preliminary data of this study without analytical and experimental proof were presented at an international conference [35].

---

[*]cbabayigit@etu.edu.tr

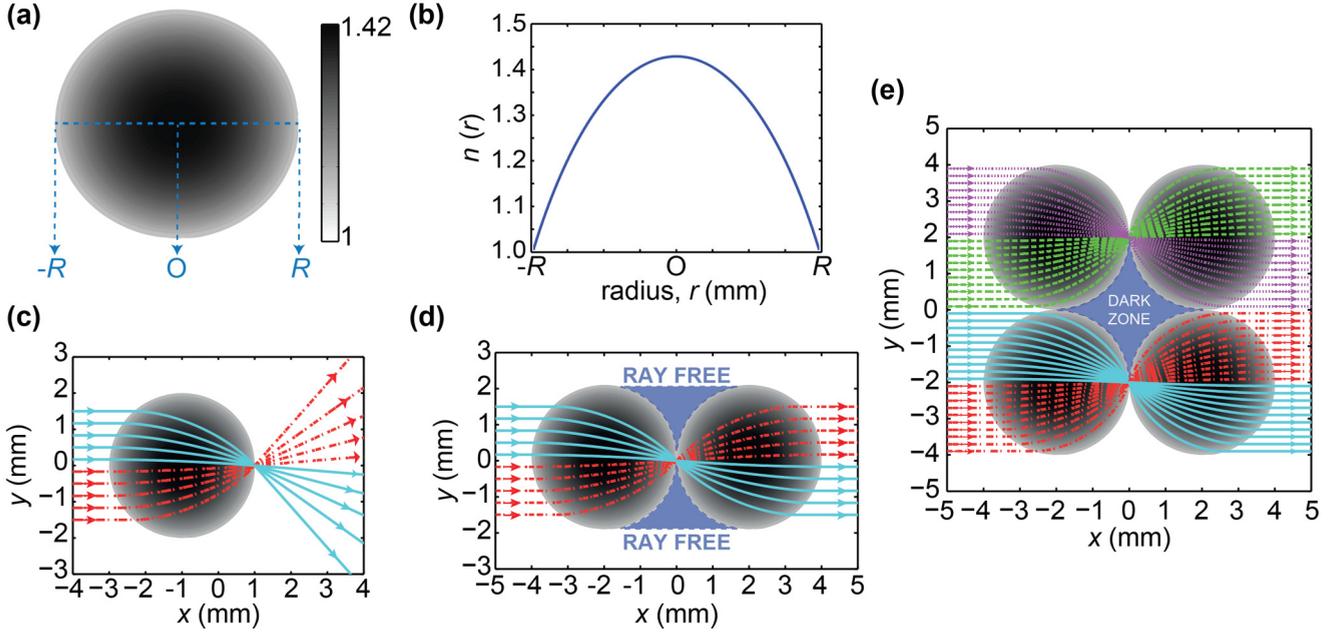

FIG. 1. (a) Schematic representation of the Luneburg lens. (b) Luneburg lens refractive index profile with respect to radius. Ray tracing of a parallel rays through (c) a single Luneburg lens, (d) double Luneburg lens system, and (e) quadruple Luneburg lens system. [Rays in (c,d) are depicted with different colors and line patterns (solid, dotted, dashed, and dash-dotted) to show image inverting (flipping upside-down) behavior of the Luneburg lens].

## II. RAY-THEORY MODEL OF THE CLOAKING BY LUNEBURG LENS

In this section, geometrical optics is used to examine the mathematical explanation of the light behavior in the Luneburg lens which is a spherical GRIN medium, where the refractive index varies radially starting from the center to the outer boundary of the lens [36]. The general index distribution of the Luneburg lens can be defined as follows:

where $n_0$ is the refractive index of the surrounding space of the lens (in our case the surrounding space is considered to be air, so $n_0 = 1$), $R$ is the radius of the lens, and $r$ is the radial polar coordinate within the lens region. In Figs. 1(a) and 1(b), schematic representation of a Luneburg lens and corresponding refractive index distribution along the polar axis [the cross section through the lens is shown by a dashed line as an inset in Fig. 1(a)] are given, respectively. Due to the radial symmetry and Luneburg lens refractive index distribution characteristic, incoming parallel rays are focused to the point at the opposite side. Also, the rays diverging from a single point located on the lens surface are collimated into parallel rays on its back surface. This special characteristic of the Luneburg lens can be analyzed by using geometrical optics based on Fermat's principle [37]. For this purpose, a quasi-two-dimensional (2D) ray solution is conducted by using the 2D medium which has the same index distribution characteristic as Luneburg lenses [38]. By combining Fermat's principle with the Lagrangian optics, the ray tracing equation for a single Luneburg lens can be represented as follows [39–41]:

$$n(r) = n_0 \sqrt{2 - \left(\frac{r}{R}\right)^2}, \quad (1)$$

$$y(x) = \frac{[2x_0 y_0 + R^2 \sin(2\theta)]x}{2x_0^2 + R^2[1 + \cos(2\theta)]} + \frac{\sqrt{2}R\sqrt{R^2[1 + \cos(2\theta)] + 2x_0^2 - 2x^2}[y_0 \cos(\theta) - x_0 \sin(\theta)]}{2x_0^2 + R^2[1 + \cos(2\theta)]}, \quad (2)$$

where $y(x)$ is a ray trajectory function with respect to position $x$, $R$ is the radius of the lens, $(x_0, y_0)$ are initial ray positions, and $\theta$ is the incidence angle of the ray. Detailed analytical derivation of Eq. (2) is provided in the Appendix. The ray trajectory equation is simplified by considering the parallel incident rays where incidence angle $\theta = 0°$:

$$y(x) = \frac{y_0\left(x_0 x + R\sqrt{R^2 + x_0^2 - x^2}\right)}{x_0^2 + R^2}. \quad (3)$$

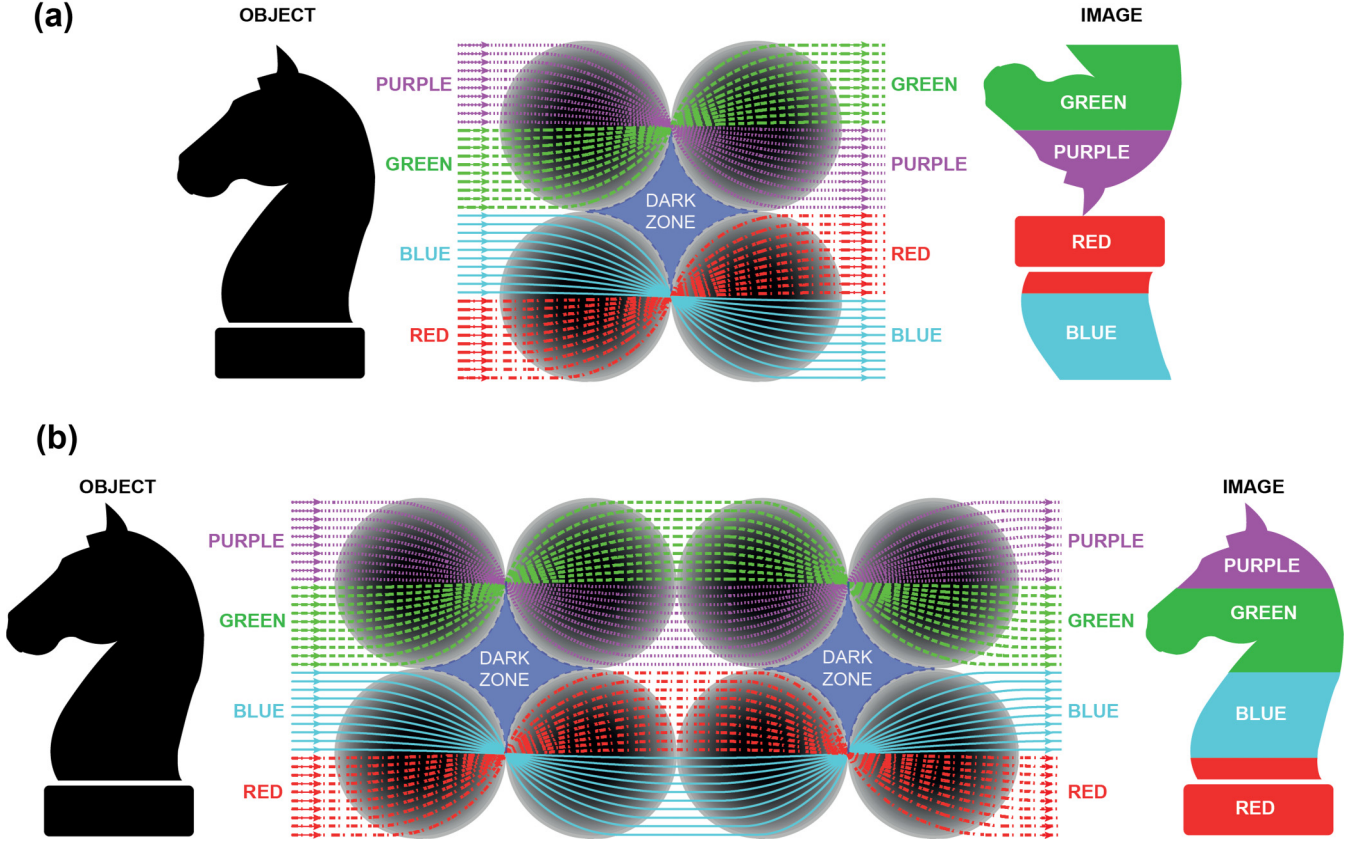

FIG. 2. (a) Schematic representation of the image formation of the arbitrary object (knight silhouette) by a quadruple Luneburg lens system where the image is separated into two parts and reversed, respectively. (b) Image correction schematic by doubling the quadruple Luneburg lens system. In (a,b), different colors and line styles (solid, dotted, dashed, and dash-dotted) are used to present the image reversing behavior of the Luneburg lens.

At the back surface of the lens, propagated light rays encounter the free space and, as a result of Snell's law, they refract with exit angles. The exit angles of the rays can be obtained from the slope information of the ray trajectory. Hence by taking the derivative of Eq. (2) with respect to $x$, propagating rays leave the lens with angles of departure according to the following equation:

$$\dot{y}(x) = \frac{R^2 \sin(2\theta) + 2x_0 y_0}{2x_0^2 + R^2[1 + \cos(2\theta)]} + \frac{2\sqrt{2}Rxy_0[\sin(\theta) - \cos(\theta)]}{\{2x_0^2 + R^2[1 + \cos(2\theta)]\}\sqrt{R^2[1 + \cos(2\theta)] - 2x^2 + 2x_0}}. \quad (4)$$

The ray trajectories are calculated by using Eqs. (2)–(4) and the corresponding ray pictures are plotted for single, double, and quadruple combinations of Luneburg lenses in Figs. 1(c)–1(e), respectively. The focusing property of the Luneburg lens, where the incoming parallel rays are focused into a single point, is proved by the solution of the ray tracing equation and the obtained result is given in Fig. 1(c). In addition, by adjoining two lenses, retransformation of the focused light into the plane wave can be observed in Fig. 1(d). In Fig. 1(d), one can see the "ray-free" regions (upside and underside regions where two lenses are connected) that are not affected by the incoming ray. From this point of view, as a next step, via the two-by-two arrangement of the Luneburg lenses, a quadruple lens system can be composed. Figure 1(e) illustrates the ray picture of the quadruple lens system. Here, one can see an isolated region from the incoming rays at the center of the quadruple lens system, which is called the "dark zone." Thus, this region can be used to electromagnetically hide an object from an incident wave.

It should be noted that due to the Luneburg lens refractive index profile, an image of the object formed by incident light is separated into two parts and reversed, respectively. Figures 2(a) and 2(b) show the flipping of the image and possible solution for the correction of the inverted or flipped image, respectively. As can be seen in Fig. 2, the isolated region defined as the dark zone is not affected by the incoming rays but when an external observer looks to the object at the back space of the quadruple Luneburg lens system, the observer detects the split and reversed image shown in Fig. 2(a). Nevertheless, the formed image is received by the observer without any distortion where all geometrical characteristics of the object are preserved. Therefore, by cascading the same quadruple lens system one can obtain a corrected image of the object as can be seen in Fig. 2(b). On the other hand, one can consider the combination of the Maxwell fisheye lenses with the Luneburg lenses as another solution to obtain a corrected image of the object where space occupancy can be reduced by 25% of overall system size. In this case, the lens system should be constructed in the following sequence of Luneburg–Maxwell fisheye–Luneburg lenses. However, when the Maxwell fisheye lenses are introduced to the system configuration, cloaking ability is achieved only in a single direction (where light is incident from left to right and right to left) direction whereas the lens system which contains only Luneburg lenses supports cloaking performance in two orthogonal directions.

In summary, the provided results in Figs. 1 and 2 exhibit the stunning potential of the quadruple Luneburg lens system that may efficiently cloak an object. In order to design a more realistic cloaking device in accordance with the cloaking concept presented in ray-theory analysis, the realistic design approach with its numerical analysis is given in the following sections.

## III. DESIGN APPROACH AND NUMERICAL ANALYSIS

In this study, the ray theory of light-ray propagation through the Luneburg lens system is presented to provide the concept of directional cloaking. Even though geometrical optics gives some insights about the operation principle of the proposed design, it is also necessary to analyze the performance of the cloaking system by conducting the FDTD method for the analysis of light matter interaction.

In general, the fabrication of the continuous GRIN media with the desired index distribution can be considered as a challenging task due to fabrication limitations. In order to overcome these difficulties, the PC structures are widely used for the approximation of continuous GRIN media via implementation of EMT. There are several methods to transform a continuous GRIN medium to a GRIN PC medium, such as appropriate arrangement of the radii of dielectric rods and air holes, adjustment of spatial distances between PC rods, and infiltration of PC air holes with different substances having different refractive indices. The main objective of these approaches is to design an inhomogeneous medium with the desired index profile by gradual change of the filling factors of elementary PC cells [42]. In the present study, in order to discretize a continuous GRIN Luneburg lens, the Maxwell Garnett EMT is employed. From the Maxwell Garnett EMT [43], the equation of the effective permittivity for transverse electric (TE) wave polarization can be expressed as follows:

$$\varepsilon_{\text{eff}} = \varepsilon_{\text{host}} + \frac{2f\varepsilon_{\text{host}}(\varepsilon_{\text{air}} - \varepsilon_{\text{host}})}{2\varepsilon_{\text{host}} + (1-f)(\varepsilon_{\text{air}} - \varepsilon_{\text{host}})}, \quad (5)$$

where $\varepsilon_{\text{host}}$ and $\varepsilon_{\text{air}}$ are the permittivity values of the host media and air holes, respectively, while $f = \pi r^2/(a^2)$ represents the dielectric filling ratio, with $r$ being the radius of the air hole. Finally, the variation formula of radii of air holes for TE polarization $r_{\text{TE}}$ can be expressed as follows:

$$r_{\text{TE}} = a\sqrt{\frac{[\varepsilon_{\text{host}} - n(x,y)^2](\varepsilon_{\text{host}} + \varepsilon_{\text{air}})}{\pi[\varepsilon_{\text{host}} + n(x,y)^2](\varepsilon_{\text{host}} - \varepsilon_{\text{air}})}}, \quad (6)$$

where $a$ is the lattice constant and $n(x, y)$ is the refractive index function of the Luneburg lens. By using (1) and (6), one can design the square lattice GRIN PC Luneburg lens for the desired host media.

In order to design the proposed cloaking concept by considering fabrication and characterization facilities, the host medium is selected as PLA material. The permittivity of the PLA material was selected as $\varepsilon_{\text{PLA}} = 2.4025$ in accordance with the Nicolson-Ross and Weir method [44] for the microwave regime. The schematic view of the designed GRIN PC Luneburg lens is presented in Fig. 3(a) with corresponding structural parameters. Moreover, the PC unit cells having air holes with maximum and minimum radii are given as insets in the same figure plot where $r_{\max} = 0.48a$ and $r_{\min} = 0.17a$, respectively. A corresponding 3D view of the refractive index profile is shown in Fig. 3(b) where gradual variation of index values occurs between 1.13 and 1.49. Here, to define the index profile, the limits of operating wavelengths should be validated in accordance with EMT. As is well known, the EMT is valid in the long-wavelength limit where the dielectric constituents can be averaged over the dielectric unit cell as an isotropic effective index. In order to define that long-wavelength limit, where application of EMT is valid, the dispersion relations of constituent PC unit cells (air holes drilled in PLA dielectric slab) are calculated by exploiting the plane-wave expansion (PWE) method [45]. The calculated dispersion diagrams of the first band in the $\Gamma X$ direction for the largest and smallest air holes' radii of PC unit cells are depicted in Fig. 3(c) where the direction of $\Gamma X$ is presented by giving the irreducible Brillouin zone. As can be deduced from this figure, the dispersion relation is nearly linear in the normalized frequency interval between $a/\lambda = 0.01$ and $a/\lambda = 0.30$. By using the slope information of the corresponding band diagrams, effective refractive index curves are extracted and depicted in Fig. 3(d). Here, effective refractive index values are nearly constant (linear) in the normalized frequency interval of $a/\lambda = 0.01$ and $a/\lambda = 0.25$. Hence, operating within this normalized frequency range ensures the validity of EMT.

The main purpose of this study is to design a cloaking device which utilizes a Luneburg lens system. The concept of the proposed cloaking approach is explained in the ray-theory part of the study. If we consider plane-wave propagation through a proposed lens system, one can observe that the light inherently bends around the dark zone because of the strong focusing effect. Hence, this zone is inherently bypassed by the light waves so that this effect can be used for hiding an object from the electromagnetic light waves. This region can be considered as an "electromagnetically hidden" region so that one can design a directional cloaking device for the concealment of an object from the electromagnetic waves, using the quadruple lens system as shown in Fig. 1(e). In this regard, the GRIN PC Luneburg lens, which is designed and presented in Fig. 3(a), can be considered as a composing part for the proposed quadruple lens system. Here, the GRIN PC Luneburg lens in Fig. 3(a) is scaled by adjusting the lattice constant as $a = 2.87$ mm in order to operate in the microwave regime. It is important to note that fabrication of the cloaking device and its experimental verification are also considered while fixing the lattice constant.

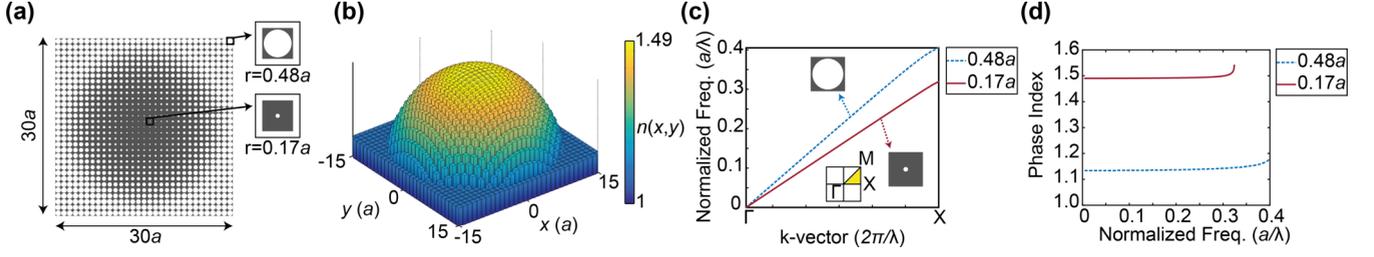

FIG. 3. (a) Schematic representation of designed Luneburg GRIN PC lens and its (b) stairstep (discrete) version of the index profile for frequency interval of $a/\lambda = 0.01$ and $a/\lambda = 0.25$. (c) The dispersion diagram of the first band along the $\Gamma X$ direction is shown. (d) Phase index curves corresponding to each dispersion band shown in (c). Note that the units of the length and frequency throughout this work are taken as $a$ and $a/\lambda$, respectively. Here, $\lambda$ is the wavelength of incident light.

Figure 4, in general, presents the conceptual design of the proposed cloaking device with its effective refractive index profile. As it can be seen from Fig. 4(a), the cloaking device is constructed by the junction of the four GRIN PC Luneburg lenses. An air hole is intentionally reamed at the center of the cloaking structure in which to place the object that is being considered for concealing. Here, the position of the air hole is arranged in a way that it overlaps with the emerging dark zone shown in Fig. 1(e). The designed cloaking device has width and length equal to 172 mm as shown in Fig. 4(b) and the slab thickness is fixed to $h = 24$ mm. In Figs. 4(a) and 4(b), the arrows indicate the direction of wave propagation. As a cloaking object, the cylindrical shaped PEC material with diameter of 36 mm and height of 24 mm is used. The corresponding 3D view of the effective refractive index profile of the proposed cloaking system is presented in Fig. 4(c) where the dark zone is defined by the dashed circle at the center of the profile.

In order to numerically analyze the cloaking performance of the proposed structure, the FDTD method is employed.

In the FDTD simulations, we used a plane-wave source to excite the cylindrical PEC object to analyze its scattering characteristics in a free-space medium. It is important to note that PEC material is a conductive material with infinite conductivity, which results in 100% reflection and 0% absorption characteristic against an incident electromagnetic field. For this reason, to clarify the cloaking performance of the proposed structure, PEC material is used through the numerical analyses. The TE polarized plane-wave source operating at a microwave frequency of 8 GHz (this frequency is obtained by transforming the normalized frequency of $a/\lambda = 0.08$ to 8 GHz) is utilized to illuminate the cloaking structure. It should be noted that, for TE polarization, the electric field components are along the $xy$ plane ($E_x$, $E_y$) and the magnetic field ($H_z$) is perpendicular to the $xy$ plane. The magnetic field ($H_z$) and phase ($\varphi$) distributions for the cases when only PEC structure was placed in free space and the cloaking structure covered the PEC object are calculated and given in Figs. 5(a) and 5(b), respectively. As can be clearly observed from magnetic field distribution in Fig. 5(a) (top), the incident plane wave is strongly scattered by the PEC object. Also, the corresponding phase distribution gives evidence of the wave front's deterioration as shown in Fig. 5(b) (top). On the other hand, when the PEC is covered by the designed cloaking structure, the scatterings are substantially suppressed where the PEC region remains isolated. Here, the magnetic field and phase profiles stay undeformed at the output of the structure as can be seen in Figs. 5(a) (bottom) and 5(b) (bottom), respectively. Thus, the plane-wave propagation nature of the incident light is mostly conserved even when a PEC material is placed inside the structure. On the other hand, impedance mismatching at the input and output faces of the cloaking structure results in reflections at these interfaces. The resulting reflections cause an optical power difference between incident and outgoing waves. Nevertheless, unwanted reflections can be suppressed by using appropriate antireflection coatings.

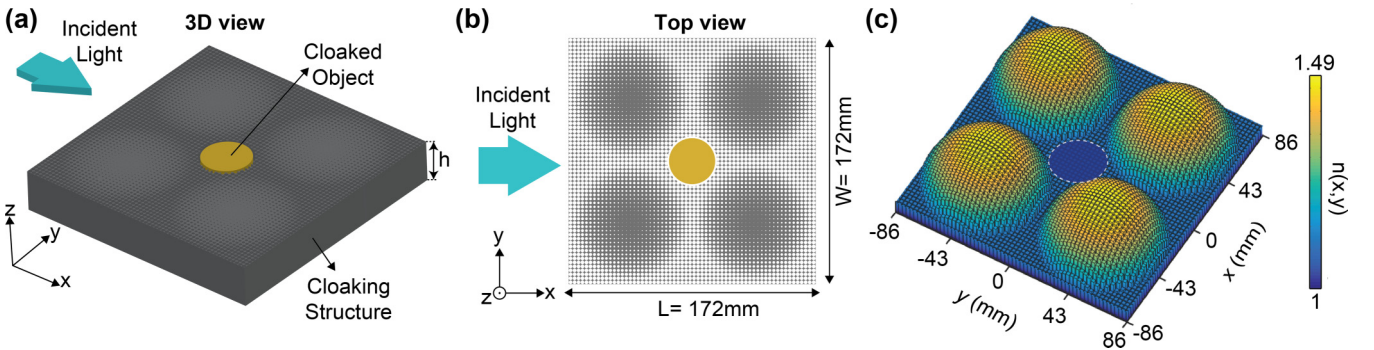

FIG. 4. (a) Three-dimensional and (b) top views of the designed cloak. (c) The stairstep (discrete) effective index profile of the proposed cloaking system.

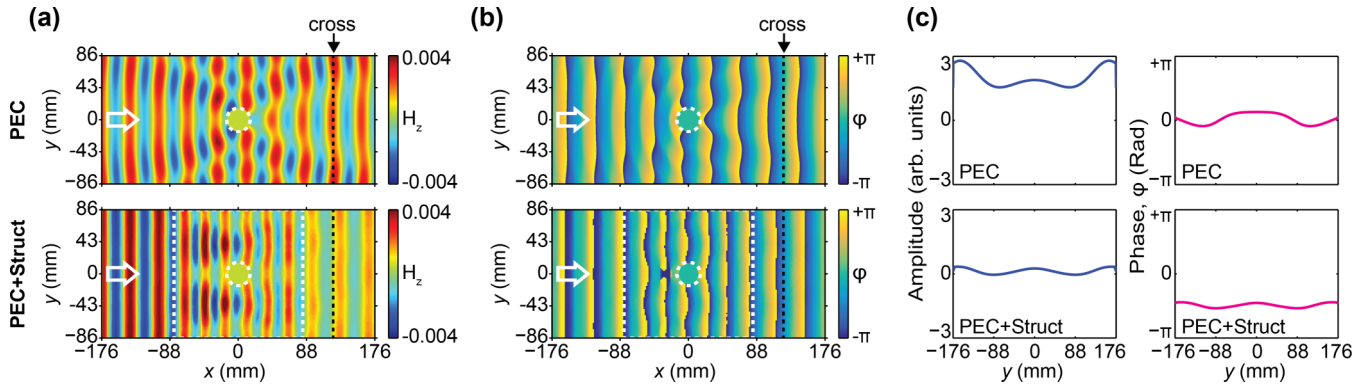

FIG. 5. The numerically calculated (a) magnetic field and (b) phase distributions are given for the cases of PEC placed in free space (top), and PEC is surrounded by cloaking structure (bottom). The arrows indicate the direction of propagation for the incident plane waves. (c) The plotted amplitude and phase profiles are calculated at the positions denoted by dashed lines in (a,b).

In order to quantitatively analyze the performance of the cloaking device, cross-sectional amplitude and phase profiles are calculated at the positions [dashed lines in Figs. 5(a) and 5(b)] and shown in Fig. 5(c). As can be seen from this figure, the PEC object highly scatters the incident wave which results in high fluctuations of cross-sectional amplitude and phase profiles. On the other hand, emerged fluctuations in the field amplitude and phase are considerably smoothed by the cloak introduced to the PEC object as shown in Fig. 5(c) (bottom). One can conclude that the decrement of fluctuations in cross-sectional profiles indicates the formation of plane waves at the output that replicates the incident one. This effect demonstrates the cloaking ability of the proposed quadruple Luneburg lens system. It should also be pointed out that all the numerical analyses, which were obtained by using the ray theory, are in accordance with the FDTD simulations.

## IV. EXPERIMENTAL VERIFICATION OF THE NUMERICALLY ANALYZED CLOAKING EFFECT IN THE MICROWAVE REGIME

The experimental verification of the cloaking concept via a quadruple Luneburg lens system is achieved by performing experiments at microwave frequencies. In this regard, the designed cloaking structure is fabricated by 3D printing of PLA material. Here, as aforementioned, the permittivity of PLA material is equal to $\varepsilon_{PLA} = 2.4025$. In order to make a realistic experiment, a cylindrical brass object was used for experiments in place of the PEC object. Here, the brass metallic object is a mixture of copper and zinc, and exhibits strong scattering characteristics at the microwave frequencies of interest. During the experimental process, we used an Agilent E5071C ENA vector network analyzer to generate and detect microwaves. The fabricated cloaking structure is excited by using a horn antenna with operating bandwidth of 6–12 GHz placed in front of it. Since the horn antenna radiates a Gaussian profiled wave, it was located at an adequate distance away from the cloaking structure to obtain plane-wave-like propagating waves. Moreover, a monopole antenna placed on a motorized stage was utilized to measure the magnetic field ($H_z$) and phase ($\varphi$) distributions at the scanning field behind the fabricated cloaking structure. In order to measure the magnetic field and phase distributions, the monopole antenna was moved by 2-mm steps along both the $x$ and $y$ axes. It should be noted that the scanning field is on a level with half of the thickness of the cloaking structure in the $z$ direction. The complete schematic representation of the experimental setup is depicted in Fig. 6(a). Furthermore, the photographic view of the fabricated cloaking structure with a cylindrical brass object placed inside it is given in Fig. 6(b) where a coin is placed for visual comparison of dimensions of the structure and brass object.

Initially, we measured the scatterings due to the brass object. Therefore, we positioned the brass object in front of the horn antenna, and then scanned the magnetic field and phase distributions behind it via the monopole antenna. Next, we placed the cloaking structure with the brass object at a sufficient distance away from the horn antenna and measured the magnetic field and phase distributions behind the cloaking structure. The corresponding magnetic field and phase distributions were measured for an incident wave with frequency of 8 GHz and represented in Figs. 6(c) and 6(d), respectively. As can be seen from the top image of Fig. 6(c), the incident light is strongly scattered and it is divided into two branches by making a shadow at the back surface of the brass object. On the other hand, the cloaking structure reduces the scatterings (almost no sign of existence of the brass object) and one can conclude that the cloaking structure has managed to equalize the magnetic field amplitude along the $y$ axis by altering the scattered fields due to the brass object. In order to give a complete picture of cloaking, the measured phase distributions are given in the top and bottom images of Fig. 6(d), respectively, for the brass object alone and the cloaking structure with the brass object inside it. Here, the brass object scatters the incident wave to have curved wave fronts behind it, while the cloaking structure reshapes the propagated fields to have plane wave fronts. The dashed lines in Figs. 6(c) and 6(d) indicate the positions of cross-sectional profiles of magnetic field and phase distributions. The corresponding cross-sectional profiles are plotted in Fig. 6(e) to demonstrate the scatterings of the brass object and the cloaking effect of the quadruple Luneburg lens system. As can be seen from the plots in Fig. 6(e), the fluctuations in the amplitude and phase profiles are reduced when the cloaking structure is introduced to cloak the brass object.

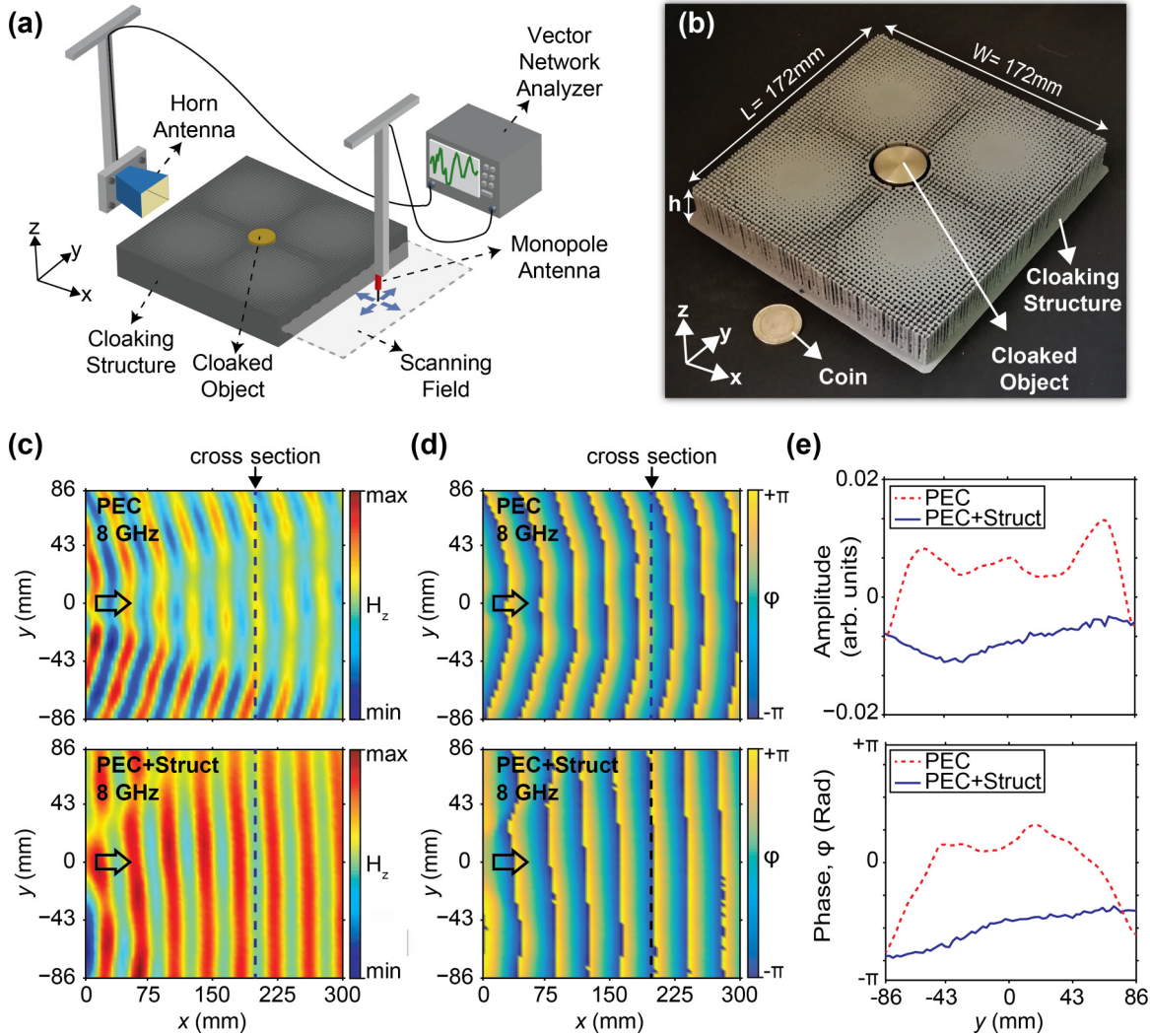

FIG. 6. (a) Schematic representation of the experimental setup. (b) The photographic view of fabricated cloaking structure and cylindrical brass cloaked object. A coin is shown for comparison of structural dimensions. (c) The measured magnetic field ($H_z$) and (d) phase ($\varphi$) distributions in the scanning field at a frequency of 8 GHz for experiment cases with the brass object alone (PEC) and the cloaking structure with the brass object inside (PEC + Struct). The arrows denote the direction of incident wave and the dashed lines indicate the positions of cross-sectional profiles. (e) The corresponding cross-sectional profiles of magnetic field amplitude and phase.

## V. CONCLUSION

In summary, the directional cloaking ability of the proposed quadruple Luneburg lens system is presented. The ray analysis of a single Luneburg lens is examined which is later adapted to the formation of double and quadruple Luneburg lenses. As a result, electromagnetically hidden space emerges in the middle of a quadruple Luneburg lens system which is an adequate region for optical cloaking of highly scattering objects such as PEC materials. In order to enable the feasibility of the refractive index distribution of a Luneburg lens, Maxwell Garnett EMT is applied by inserting air holes with gradually varying radii on a dielectric slab which results in a GRIN PC Luneburg lens. The host dielectric material is selected as PLA with permittivity of $\varepsilon_{PLA} = 2.4025$ and the EMT approximations are performed via the PWE method which is discussed in detail. Further, cloaking of a cylindrical PEC object by the quadruple GRIN PC Luneburg lens system is analyzed by employing the FDTD method where the system is excited by a TE polarized plane-wave source. According to the numerical results, the designed system is able to cloak the object and reproduce the plane-wave aspects of the incident wave at the output area. Additionally, the proposed Luneburg lens system is fabricated via 3D printing technology and microwave experiments are performed at a frequency of around 8 GHz for the verification of analytical solutions and numerical calculations of the directional cloaking effect. During the experiments, brass alloy is preferred in substitution for the PEC material and it is observed that the printed-out lens system managed to suppress the scatterings due to the highly scattering brass object. As a result, we designed, fabricated, and both numerically and experimentally validated the proposed cloaking concept of the quadruple Luneburg lens system. The designed structure can be a good solution candidate for optical cloaking from the incident light in a defined propagation direction. One can increase the numbers of Luneburg lenses to realize even larger arrays to hide multiple objects.

## ACKNOWLEDGMENTS

This work was supported by the Scientific and Technological Research Council of Turkey (TUBITAK) under Project No. 116F182. H.K. acknowledges partial support from the Turkish Academy of Sciences.

## APPENDIX: DERIVATION OF RAY TRAJECTORY EQUATION

The optical path length of a ray in an inhomogeneous medium between points A and B can be defined as follows [36,37]:

$$\text{OPL} = \int_A^B n(r)ds, \quad (A1)$$

where $n(r)$ corresponds to the refractive index function, which varies with position, and the differential length is expressed as $ds = \sqrt{dr^2 + r^2 d\varphi^2}$ in polar coordinates. Since the refractive index distribution of the Luneburg lens is a function of $r$, so differential length $ds$ can be rewritten as $ds = \sqrt{1 + r^2(\frac{d\varphi}{dr})^2} dr$. Let $\frac{d\varphi}{dr} = \dot{\varphi}$, then Eq. (A1) becomes

$$\text{OPL} = \int_A^B n(r)\sqrt{1 + r^2 \dot{\varphi}^2} dr. \quad (A2)$$

Here, the shortest path that is followed by a light ray according to Fermat's principle can be obtained by the minimization of the integral of (A2). In this regard, to obtain the derivative of (A2), the Euler-Lagrange equation can be used where the Lagrangian is $L(\varphi, \dot{\varphi}, r) = n(r)\sqrt{1 + r^2 \dot{\varphi}^2}$ [38,41]:

$$\frac{d}{dr}\frac{\partial L}{\partial \dot{\varphi}} = \frac{\partial L}{\partial \varphi}. \quad (A3)$$

Taking into consideration that the structure is invariant in the $\varphi$ direction, the derivative of $\partial L/\partial \varphi$ becomes equal to zero.

Then, the left-hand side of (A3), $\partial L/\partial \dot{\varphi}$, should be a constant value, since $\frac{d}{dr}\frac{\partial L}{\partial \dot{\varphi}} = 0$. Taking the derivative of the Lagrangian equation with respect to the $\dot{\varphi}$ and equalizing it to constant $C_1$, we obtain

$$\frac{\partial L}{\partial \dot{\varphi}} = \frac{n(r)r^2}{\sqrt{1 + r^2 \dot{\varphi}^2}} \dot{\varphi} = C_1. \quad (A4)$$

Next, the nonlinear differential equation (A4) can be rewritten by replacing $\dot{\varphi}$ with $d\varphi/dr$:

$$\frac{n(r)r^2}{\sqrt{1 + r^2 \left(\frac{d\varphi}{dr}\right)^2}} \frac{d\varphi}{dr} = C_1. \quad (A5)$$

Then, as a step to reach ray trajectory $r(\varphi)$, one should solve (A5) for $d\varphi$ and integrate both sides as follows:

$$\varphi = \frac{1}{2}\sin^{-1}\left[\frac{r^2 - \frac{C_1^2 R^2}{n(r)^2}}{r^2 \sqrt{1 - \frac{C_1^2}{n(r)^2}}}\right] - \beta, \quad (A6)$$

where $R$ is the radius of the lens. Next, by solving (A6) with respect to $r$, the ray trajectory equation can be found as follows:

$$r(\varphi) = \frac{C_2 R}{\sqrt{1 - \sqrt{1 - C_2^2} \sin[2(\varphi + \beta)]}}, \quad (A7)$$

where $C_2 = C_1^2/n(r)^2$ and $\beta$ are constants. Since we are dealing with ray propagation in Cartesian coordinates, it is logical to transfer the ray trajectory formula (A1) to the Cartesian coordinate [36,39]:

$$[1 - T\sin(2\beta)]x^2 + [1 + T\sin(2\beta)]y^2 - 2T\cos(2\beta)xy + (T^2 - 1)R^2 = 0. \quad (A8)$$

$T$ and $\beta$ are constant numbers, and to find them the boundary conditions which are related with the initial position and the initial incident angle should be used. The first boundary condition depends on the initial position of the ray, where it enters the lens. Let us place the center of the lens to the position $[x = 0, y = 0]$. If the angle of incidence is $\theta$, then the initial ray position is expressed as $[x_0 = -R\cos(\theta), y_0 = -R\sin(\theta)]$. By replacing $x$ and $y$ with the position of the initial point in (A2), one can solve (A2) with respect to constant number $T$ as follows:

$$T = \sin(2\beta + 2\theta). \quad (A9)$$

The second boundary condition comes from the $dy/dx = \tan(\theta)$ relation and by taking the derivative of (8) with respect to $x$, and then solving it for $T$ by setting $x = x_0$ and $y = y_0$:

$$T = \frac{x_0 + y_0 \tan(\theta)}{\tan(\theta)[x_0 \cos(2\beta) - y_0 \sin(2\beta)] + [x_0 \sin(2\beta) + y_0 \cos(2\beta)]}. \quad (A10)$$

Here, to find the analytical expression for $\beta$, (A3) and (A4) are combined and solved for $\beta$ as follows:

$$\beta = \frac{1}{2}\left[\tan^{-1}\left(\frac{x_0}{y_0}\right) - \theta\right]. \quad (A11)$$

Then, if $\beta$ is inserted into (A3), both the $T$ and $\beta$ constants will be represented in terms of the initial positions $[x_0, y_0]$ and the incidence angle $\theta$:

$$T = \sin\left[\tan^{-1}\left(\frac{x_0}{y_0}\right) + \theta\right]. \tag{A12}$$

After finding the constants $T$ and $\beta$, (A2) can be solved for $y$ by using the quadric formula as follows:

$$y = \frac{T\cos(2\beta)x}{1 - T\sin(2\beta)} + \frac{\sqrt{[T\cos(2\beta)xy]^2 - [1 - T\sin(2\beta)]\{(T^2 - 1)R^2 + [1 + T\sin(2\beta)]x^2\}}}{1 + T\sin(2\beta)}. \tag{A13}$$

To simplify the representation of (A7), let us divide it into two components as follows:

$$A = \frac{T\cos(2\beta)x}{1 - T\sin(2\beta)}, \tag{A14}$$

and

$$B = \frac{\sqrt{[T\cos(2\beta)xy]^2 - [1 - T\sin(2\beta)]\{(T^2 - 1)R^2 + [1 + T\sin(2\beta)]x^2\}}}{1 + T\sin(2\beta)}. \tag{A15}$$

Next, solve them separately, and then merge them to obtain the final result. In this regard, firstly, we substitute the formulas of constants $T$ and $\beta$ to the term $A$ as follows:

$$A = \frac{T\cos(2\beta)x}{1 - T\sin(2\beta)} = \frac{x\sin[\tan^{-1}(x_0/y_0) + \theta]\cos[\tan^{-1}(x_0/y_0) - \theta]}{1 + \sin[\tan^{-1}(x_0/y_0) + \theta]\sin[\tan^{-1}(x_0/y_0) - \theta]}. \tag{A16}$$

Then, with the help of trigonometric identities, (A10) can be further simplified to the following equation:

$$A = \frac{x\sin[\tan^{-1}(x_0/y_0)]\cos[\tan^{-1}(x_0/y_0)]}{1 + 0.5\{[1 + \cos(2\theta)] - 2\cos^2[\tan^{-1}(x_0/y_0)]\}} + \frac{0.5\sin(2\theta)x}{1 + 0.5\{[1 + \cos(2\theta)] - 2\cos^2[\tan^{-1}(x_0/y_0)]\}}. \tag{A17}$$

As a next step, by setting $\alpha = \tan^{-1}(x_0/y_0)$ and using the circle equation, one can obtain the definitions $\sin(\alpha) = (x_0/R)$ and $\cos(\alpha) = (y_0/R)$. In this sense, (A11) can be rewritten as follows:

$$A = \frac{[\sin(\alpha)\cos(\alpha) + 0.5\sin(2\theta)]x}{1 + 0.5\{[1 + \cos(2\theta)] - 2\cos^2(\alpha)\}} = \frac{x[(x_0/R)(y_0/R) + 0.5\sin(2\theta)]x}{1 + 0.5\{[1 + \cos(2\theta)] - 2(y_0/R)^2\}} = \frac{[2x_0y_0 + R^2\sin(2\theta)]x}{2x_0^2 + R^2[1 + \cos(2\theta)]}. \tag{A18}$$

Equation (A12) is the last version of the first term $A$. The next goal is simplifying the second term $B$. For that purpose, (A9) can be rewritten as indicated below:

$$B = \frac{\sqrt{[T\cos(2\beta)xy]^2 - [1 - T\sin(2\beta)]\{(T^2 - 1)R^2 + [1 + T\sin(2\beta)]x^2\}}}{1 + T\sin(2\beta)}$$

$$= \frac{\sqrt{(T^2 - 1)x^2 + (T^2 - 1)[-T\sin(2\beta) + 1]R^2}}{1 + T\sin(2\beta)}$$

$$= \frac{\sqrt{(T^2 - 1)\{x^2 - R^2[1 + T\sin(2\beta)]\}}}{1 + T\sin(2\beta)}. \tag{A19}$$

There are two important expressions in (A13), which are $(T^2-1)$ and $[1 + T\sin(2\beta)]$. By substituting the formulas of constants $T$ and $\beta$ to these expressions, the $B$ term will be written in terms of the initial position and the initial incident angle. For that purpose, the following operations were carried out with the mentioned expressions:

$$T^2 - 1 = \sin[\tan^{-1}(x_0/y_0) + \theta]^2 - 1 = \sin(\alpha + \theta)^2 - 1 = \frac{1 - \cos(2\alpha + 2\theta)}{2} - 1$$

$$= \frac{-1 - [2\cos^2(\alpha) - 1]\cos(2\theta) - 2\sin(\alpha)\cos(\alpha)\sin(2\theta)}{2}$$

$$= \frac{-R^2 - (2y_0^2 - R^2)\cos(2\theta) + 2x_0y_0\sin(2\theta)}{2R^2}, \tag{A20}$$

$$1 + T\sin(2\beta) = 1 + \sin[\tan^{-1}(x_0/y_0) + \theta]\sin[\tan^{-1}(x_0/y_0) - \theta] = 1 + \sin(\alpha + \theta)\sin(\alpha - \theta)$$

$$= \frac{3 + \cos(2\theta) - 2\cos^2(\alpha)}{2} = \frac{[1 + \cos(2\theta)]R^2 + 2x_0^2}{2R^2}. \tag{A21}$$

Then, by combining (A13), (A14), and (A15), the final version of the $B$ term can be obtained in the simplest way as follows:

$$B = \frac{\sqrt{2}R[y_0 \cos(\theta) - x_0 \sin(\theta)]\sqrt{R^2[1 + \cos(2\theta)] + 2x_0^2 - 2x^2}}{2x_0^2 + R^2[1 + \cos(2\theta)]}. \tag{A22}$$

Here, the simplified versions of $A$ and $B$ are attained in (A12) and (A17), respectively. Finally, by merging (A12) and (A17), we obtain the ray trajectory equation in the Cartesian coordinate system as follows:

$$y(x) = \frac{[2x_0 y_0 + R^2 \sin(2\theta)]x}{2x_0^2 + R^2[1 + \cos(2\theta)]} + \frac{\sqrt{2}Ry_0 \cos(\theta)\sqrt{R^2[1 + \cos(2\theta)] + 2x_0^2 - 2x^2}}{2x_0^2 + R^2[1 + \cos(2\theta)]}$$
$$- \frac{\sqrt{2}Rx_0 \sin(\theta)\sqrt{R^2[1 + \cos(2\theta)] + 2x_0^2 - 2x^2}}{2x_0^2 + R^2[1 + \cos(2\theta)]}, \tag{A23}$$

where $y(x)$ is a ray trajectory function with respect to position $x$, $R$ is the radius of the lens, $(x_0, y_0)$ are initial ray positions, and $\theta$ is the incidence angle of the ray. Moreover, the exit angle, which is the angle at which the ray exits the lens, should be computed for the complete ray analyses. In order to determine the exit angle, the derivative of (A17) should be taken with respect to $x$ as follows:

$$\dot{y}(x) = \frac{R^2 \sin(2\theta) + 2x_0 y_0}{2x_0^2 + R^2[1 + \cos(2\theta)]} + \frac{2\sqrt{2}Rxy_0[\sin(\theta) - \cos(\theta)]}{\{2x_0^2 + R^2[1 + \cos(2\theta)]\}\sqrt{R^2[1 + \cos(2\theta)] - 2x^2 + 2x_0}}. \tag{A24}$$

Equations (A17) and (A18) can be used to investigate multiple configurations of Luneburg lenses. In this regard, the ray trajectories that are obtained analytically by using (A17) and (A18) for distinct cases are represented in Fig. 7.

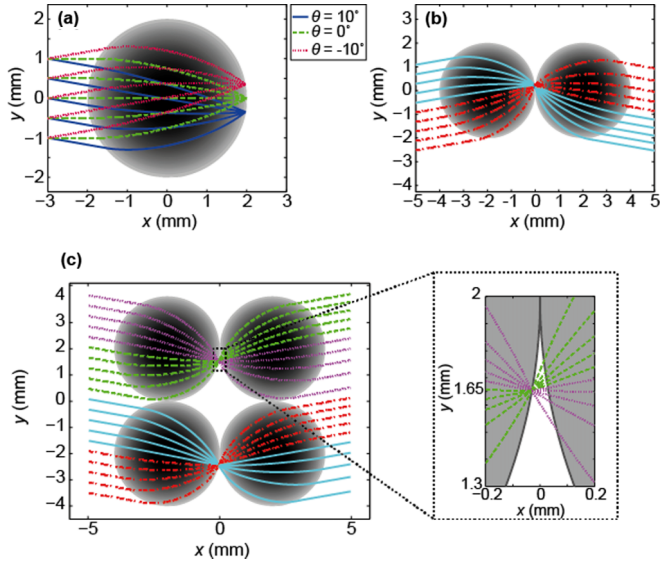

FIG. 7. Ray tracing of parallel rays through (a) a single Luneburg lens with incidence angles $\theta = 10°$, $0°$, and $-10°$; (b) double Luneburg lens system with an incidence angle $\theta = -10°$; (c) quadruple Luneburg lens system with an incidence angle $\theta = 10°$ where the lens junction region is enlarged and given as an inset.

In Fig. 7(a), the focusing characteristic of the Luneburg lens under the different incidence angles is observed. From the figure, one can conclude that, depending on the incidence angle of the rays, the focal point shifts on the surface of the lens. Moreover, by adjoining two, four, or more lenses, specific Luneburg lens systems can be designed. Here, in Figs. 7(b) and 7(c) ray trajectories for double and quadruple lens systems are presented where the incidence angles are chosen as $-10°$ and $10°$, respectively. As can be seen from Figs. 7(b) and 7(c), even though light is incident with an angle different from normal, the ray-free zone has emerged.

Since incoming parallel rays are focused into a point at the exact opposite surface of the Luneburg lens, the focal point is no longer on the tangent point of the adjacent lenses under the oblique incidence angle of $10°$ as can be seen in Fig. 7(c). Here, the enlarged junction region shows that off-axis focused light rays at the surface (shifted downward with respect to the junction position of two Luneburg lenses) first diverged into the free space and, after propagating small distance, they entered the second lens at nonuniformly spaced positions. In this case, the spacing of the light rays emanating from the quadruple Luneburg lens becomes nonuniform under the oblique incidence case. Hence, the obliqueness of the incident wave may cause distortion of the image. On the other hand, by applying the same image correction concept (two cascaded quadruple lens systems) which is discussed in Sec. II [see Fig. 2(b)], one can regenerate the distorted image outside of the lens system for small incidence angles.

In summary, for certain initial conditions, the analyses of several configurations of Luneburg lenses can be examined with the help of the ray trajectory equation given in (A17).